\documentclass[conference]{IEEEtran}
\IEEEoverridecommandlockouts
\usepackage{cite}
\usepackage{amsmath,amssymb,amsfonts}
\usepackage{algorithmic}
\usepackage{graphicx}
\usepackage{textcomp}
\usepackage{xcolor}
\def\BibTeX{{\rm B\kern-.05em{\sc i\kern-.025em b}\kern-.08em
    T\kern-.1667em\lower.7ex\hbox{E}\kern-.125emX}}

\usepackage{multirow}
\usepackage{makecell}
\usepackage{booktabs}
\usepackage{mathtools}

\begin{document}


\title{Removal of Ocular Artifacts in EEG Using Deep Learning\\

}

\author{
\IEEEauthorblockN{Mehmet Akif Ozdemir}
\IEEEauthorblockA{\textit{Dept. of Biomedical Eng.} \\
\textit{Izmir Katip Celebi University}\\
Izmir, Türkiye \\
makif.ozdemir@ikcu.edu.tr}
\and
\IEEEauthorblockN{Sumeyye Kizilisik}
\IEEEauthorblockA{\textit{Dept. of Biomedical Eng.} \\
\textit{Izmir Katip Celebi University}\\
Izmir, Türkiye \\
sumeyye.kizilisik@gmail.com}
\and
\IEEEauthorblockN{Onan Guren}
\IEEEauthorblockA{\textit{Dept. of Biomedical Eng.} \\
\textit{Izmir Katip Celebi University}\\
Izmir, Türkiye \\
onan.guren@ikcu.edu.tr}
}


\maketitle



\begin{abstract}
EEG signals are complex and low-frequency signals. Therefore, they are easily influenced by external factors. EEG artifact removal is crucial in neuroscience because artifacts have a significant impact on the results of EEG analysis. The removal of ocular artifacts is the most challenging among these artifacts. In this study, a novel ocular artifact removal method is presented by developing bidirectional long-short term memory (BiLSTM)-based deep learning (DL) models. We created a benchmarking dataset to train and test proposed DL models by combining the EEGdenoiseNet and DEAP datasets. We also augmented the data by contaminating ground-truth clean EEG signals with EOG at various SNR levels. The BiLSTM network is then fed to features extracted from augmented signals using highly-localized time-frequency (TF) coefficients obtained by wavelet synchrosqueezed transform (WSST). We also compare the WSST-based DL model results with traditional TF analysis (TFA) methods namely short-time Fourier transformation (STFT) and continuous wavelet transform (CWT) as well as augmented raw signals. The best average MSE value of 0.3066 was obtained by the first time-proposed BiLSTM-based WSST-Net model. Our results demonstrated the WSST-Net model significantly improves artifact removal performance compared to traditional TF and raw signal methods. Also, the proposed EOG removal approach reveals that it outperforms many conventional and DL-based ocular artifact removal methods in the literature.
\end{abstract}


\section{Introduction}

Electroencephalography (EEG) is a beneficial approach for identifying brain signals corresponding to various conditions on the scalp surface. It is a valuable neuroimaging tool for measuring evoked potentials (EPs), which are voltage fluctuations caused by evoked neuronal activity, or for use in applications like brain-computer interfaces (BCI). Due to the low frequency of brain signals, various undesirable signals may interfere with accurate measurements \cite{teplan2002fundamentals}. Due to the multilayered skull, EEG recordings are contaminated with noise from numerous sources such as eye movements, instrumental interferences, muscle activity, and cardiac activity during measurement. In event-related potential (ERP) analysis, it is especially difficult to distinguish. This incident leads to misinterpretations and inconsistencies. Therefore, EEG signals should contain only brain activity \cite{mannan2018identification}.

Artifact removal is fundamental in clinical and practical research and it is essential to have knowledge of the general properties and types of these artifacts. Non-physiological artifacts can be minimized with proper setup. Nonetheless, physiological artifact removal necessitates the use of an appropriate complex algorithm \cite{sweeney2012artifact}. Ocular artifacts are the most common artifacts that may observe frequently in the EEG signal. Eye blinks are limited to the frontal region. It occurs as a result of a potential difference between the cornea and retina. The amplitude range is 50-200 mV \cite{ranjan2021ocular}. The peaks of ocular waves are sharper than those of cerebral activities. Repetitive eye blinking, eyelid movements, side gaze, slow eye movement, and wandering eye movements are all common eye-related artifacts\cite{ccinar2017novel}. Especially horizontal and vertical eye movements greatly affect EEG signals. Therefore the removal of these artifacts essential impact on brain-related neuroscience studies \cite{minguillon2017trendseeg}.

Several techniques are proposed to remove ocular artifacts in the contaminated EEG signal. The conventional techniques used in general for the removal of ocular artifacts include filtering, regression, the wavelet transform method, and Blind Source Separation (BSS). Independent component analysis (ICA) is one of the most common artifact removal algorithms and does not need a reference channel, unlike other methods \cite{gorjan2022removaleeg}. In particular, these methods may be compared on the basis of the reference channel, process automation, and adaptive capacity in the online environment. Although there are hybrid methods that combine several methods to reduce these drawbacks, there is no standard conventional method for the removal of ocular artifacts.

Apart from conventional methods, with the widespread usage of machine learning (ML),various studies in the field of removing EEG artifacts have been reported. Support Vector Machines (SVM) and Neural Networks (NN) are the most commonly used techniques for processing EEG artifacts \cite{barua2014review}. ICA and SVM were used together in many studies. Lawhern et al. developed a MATLAB Toolbox that uses SVM classification of EEG signals. This toolbox extracts and characterizes features from EEG signals using an autoregressive model which successfully distinguishes synthetic data \cite{lawhern2013detect}. Besides, there are many ML studies such as fuzzy inference system \cite{vijila2007artifacts}, Bayesian model \cite{schetinin2007bayesian}, and genetic programming \cite{fairley2010hybrid}. Although ICA is a frequently used method, it requires expert observation for artifact identification. In this respect, ML methods may be considered more successful than conventional methods in the manner of time consumption and performance. However, the feature vector must be manually extracted before constructing an ML model. These procedures time consuming and require a specialist's opinion. As a consequence, without pre-processing and professional assistance, these techniques cannot process raw data. Deep Learning (DL) has made significant progress in resolving the long-standing problems in ML research. DL methods obtain the relevant information using the representations that have been created in different layers while processing the data.

Admitting DL has limited work in the field of EEG noise removal, there are significant DL models that have demonstrated successful results. Nguyen et al. \cite{nguyen2012eog19}, performed a Deep Neural Network (DNN) based study using a wavelet neural network (WNN). It was reported to be more efficient and useful than ICA because it enables a fully automated online algorithm. Yang et al. \cite{yang2018automatic20}, conducted a DL-based study to eliminate the reference channel disadvantages by proposing a two-stage method. This study, which used the stacked sparse autoencoder (SSAE) as a DL architecture, achieved significant success compared to various conventional algorithms. Sun et al. \cite{sun2020novel21}, proposed a raw waveform-based one-dimensional Residual Convolutional Neural Networks (1D-ResCNN) model. The end-to-end method was used to match noisy signals to clean signals. This method has been shown to perform well compared to ICA. U-net model, a DL method commonly used for image segmentation, was used for ocular artifact removal by Mashhadi et al. \cite{mashhadi2020deep23}. In another study, a method based on deep convolutional neural networks (DCNN) was developed to remove artifacts quickly and automatically \cite{lopes2021automatic24}. Zhang et al.\cite{zhang2021eegdenoisenet26}, utilized 4 different neural networks and compare the results with conventional methods. As a result, they demonstrated that DL models outperform conventional methods and presented a benchmark EEG dataset that may be used for DL algorithms. As a result, when DL models are investigated in the literature, there are studies that may compete with ICA, which is a common approach due to its single-channel artifact removal performance. Furthermore, there are studies that demonstrate successful results in multi-channel signals. These studies are significant, especially in terms of time savings. However, they require complex pre-processing methods and their success are limited.

In this study, an extended dataset is trained using a DL model based on a novel BiLSTM-regression architecture to remove ocular artifacts in EEG. A wavelet synchrosqueezed transform (WSST)-based method is presented for extracting distinct features from contaminated EEG signals, for the first time. Also, the presented method's performance is benchmarked with traditional TF analysis (TFA) methods namely short-time Fourier transform (STFT) and continuous wavelet transform (CWT) as well as a raw signal-based approach.

\vspace{-2mm}

\section{Materials and Methods}

\subsection{Dataset}

Two distinct datasets, EEGdenoiseNet \cite{zhang2021eegdenoisenet26} and DEAP \cite{koelstra2011deap}, were used in this study. EEGdenoiseNet is a dataset composed of 4514 clean EEG segments, 3400 ocular artifact segments, and 5598 muscle artifact segments derived from publicly available data sources for use in DL models. Increasing the amount of data is one method suggested for improving the performance of DL models \cite{makifAbnormal2021}. The DEAP dataset is adopted to improve the model's performance by increasing the amount of data. DEAP consists of 32 subjects' EEG recordings collected while they watching 40 music videos. 1650 clean EEG segments and 1600 EOG segments were extracted and contributed to the EEGdenoiseNet using 2-second arbitrary selections from the DEAP dataset. In the study, a total of 6164 clean EEG segments and 5000 EOG segments were utilized. All signal segments have a sampling rate of 256 Hz.

\subsection{Pre-processing}

The data in EEGdenoiseNet has already been pre-processed. In the DEAP dataset, raw data was segmented into 2 s-parts and processed following closely the pre-processing procedures in EEGdenoiseNet \cite{zhang2021eegdenoisenet26}. The DEAP data was analyzed using EEGLAB toolbox \cite{DELORME20049}. For generating ground-truth clean EEG segments, a 1-80 Hz band-pass filter and the \textit{CleanLine} function at 50 Hz were utilized. The signals were then resampled to 256 Hz and ICA was applied. Only high-quality brain-labeled ICA components were selected and as a result, a total of 1650 clean EEG segments with 2 s-long sizes were generated. For EOG segments, horizontal and vertical EOG measurements were considered. Then, a band-pass filter with a frequency range of 0.3 to 10 Hz and a 50 Hz notch filter was performed to filter the EOG segments. Similarly, EOG segments were resampled to 256 Hz. Eventually, sharply amplified waves from these recordings were adopted to separate a total of 1600 EOG segments with 2 s-long sizes.

\subsection{Semi-synthetic data augmentation}

\begin{figure}[htbp]
\vspace{-2mm}
\centering
\centerline{\includegraphics[width=0.8\columnwidth]{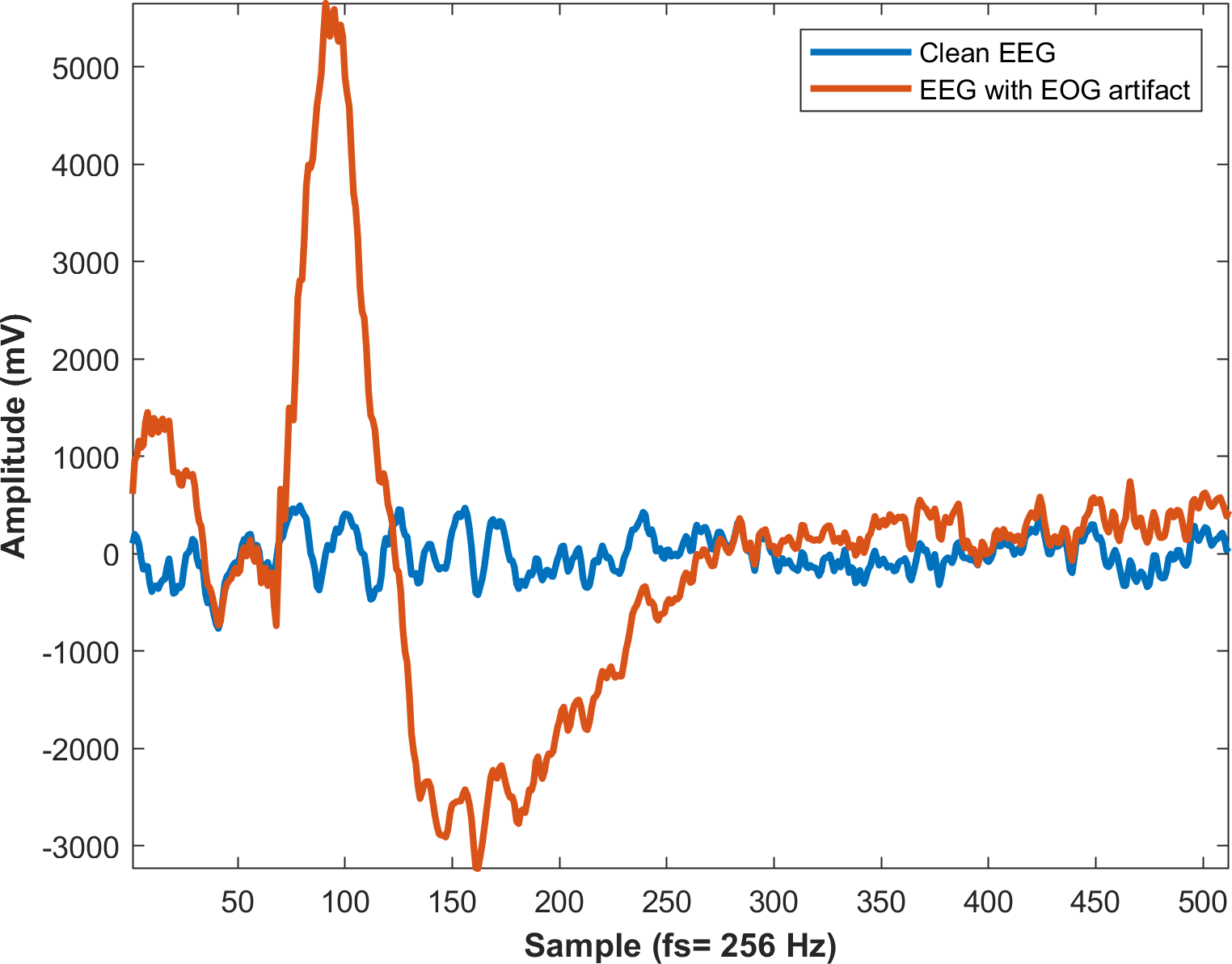}}
\caption{An example of ground truth EEG segment and the corresponding contaminated EEG segment generated at -10 dB SNR.}
\label{fig1}
\vspace{-2mm}
\end{figure}

The signal-to-noise ratio (SNR) is a measurement that compares the actual signal information with the noised signal. In this study, noisy EEG segments were generated at different SNR levels (-10 to 4 dB). Contaminated signals were generated by linearly combining clean EEG segments with EOG artifact segments using equation (\ref{equ1}). This function combines ground-truth EEG and EOG signals to create noisy EEG segments with varying SNR values. The $\lambda$ parameter may be changed to control the artifact strength and obtain a specific SNR value.
\begin{equation}
noisyEEG=EEG + \lambda.EOG
\label{equ1}
\end{equation}
The SNR levels of the contaminated EEG segment can be determined by changing the $\lambda$ by using equation (\ref{equ2}).
\begin{equation}
SNR=10\cdot log{\frac{RMS (EEG)}{RMS(\lambda.EOG)}}
\label{equ2}
\end{equation}
The root mean square (RMS) value equation is expressed in \cite{zhang2021eegdenoisenet26}. The lower $\lambda$ value represents a higher SNR value and presents a low noisy EEG signal. Likewise, a lower SNR value and higher $\lambda$ value represents a high noise. As previously reported, the SNR value of EEG contaminated with ocular artifacts ranges from -7dB to 2dB \cite{wang2015removal}. In this study, the SNR range of -10dB to 4dB was extended to examine the ocular artifact removal performance over a wider range of SNR levels. As a result, 75,000 semi-synthetic EEG segments were generated at different SNRs. Fig. \ref{fig1} illustrates an example of ground truth clean EEG segment and generated semi-synthetic contaminated EEG segments. 

\vspace{-3mm}

\subsection{Feature extraction}

The ability to use original raw signals in DL models is a significant advantage. In this context, raw EEG segments were fed to an BiLSTM-based DL architecture called RAW-Net. On the other hand, using distinct features extracted from raw data is a common approach for improving the performance of DL models. Thus, the architecture may learn more distinct characteristics of the data. With this regard, the TF characteristic of raw EEG data was extracted by utilizing STFT, CWT, and WSST, separately. STFT is a popular signal processing technique for analyzing non-stationary signals. The signal is divided into smaller segments that assumed stationary using a windowing function \cite{ozdemir2021EEG}. To improve model performance STFT features are utilized to feed the same BiLSTM structure adapted to RAW-Net architecture called STFT-Net. Similarly, CWT coefficients of the raw signals were utilized to feed the same architecture called CWT-Net. CWT is a TF method for analyzing  and displaying signal characteristics that vary with time and scale \cite{sadowsky1996investigation}. CWT adjusts the length of the main wavelet to achieve TF localization differs from STFT. CWT precisely analyzes low frequencies with long wavelets and provides high-time localization with short wavelets \cite{vetterli1991wavelets}. However, it provides limited information about frequency content. The use of a fixed-time window results in spectral smearing and leakage. Hence, TF coefficients may suffer some artifacts \cite{oppenheim1998spectrum}. Although STFT and CWT are the most popular TFA methods, they have a disadvantage that limits TF  resolutions. The synchrosqueezing transform (SST) is a TF analysis method that has been proposed to improve TF resolution. The WSST is an extension of the CWT technique and reallocates signal energies in the frequency domain while preserving time resolution \cite{kumar2020improved}. The reallocating procedure eliminates the mother wavelet spread effect. Thus, WSST can provide well TF resolutions and compensates disadvantages of traditional methods. A final model called WSST-Net is structured in the same BiLSTM architecture to improve performance. It should be noted that BiLSTM architectures can only be fed real variables and TF coefficients may contain imaginary parts. Therefore, only the real part of TF coefficients is considered for proposed BiLSTM-regression models. Additionally, normalizing the input data improves the regression network's performance when the data distribution is variable. A simple normalization technique was accomplished by subtracting the mean of the input EEG signals and dividing the result by the standard deviation prior to training the BiLSTM models.

\subsection{BiLSTM architecture}

Long short-term memory (LSTM) architectures subtypes of recurrent neural networks (RNN) and effectively solves long-term dependency problems in sequence prediction. Unlike a unidirectional LSTM, BiLSTM learns bidirectional dependencies in sequence data. Hence, the given input data is processed in both directions and one works in a reverse direction of input data. This additional LSTM layer is capable to learn the reversed sequences and eventually joins other LSTM output sequences. BiLSTM is an effective architecture for training the sequential dependencies between input signal segments in both directions of time series \cite{rahman2022extended}. 


Considering the ability to learn distinct features of time series, a novel and effective seven-layer BiLSTM-based DL architecture is presented to remove ocular artifacts from contaminated EEG signals, for the first time. The proposed BiLSTM architecture consists of an input sequence layer, the first BiLSTM layer, a dropout layer, the second BiLSTM layer, the final dropout layer, a fully connected layer, and a regression layer, respectively. While the first BiLSTM layer works on given sequences, the second layer learns from reversed data. Two different dropout layers were added to prevent overfitting with dropout rates of 30\% and 20\%, respectively. ReLu activation function was adapted with a fully connected layer. Lastly, a regression layer was composed of outputs of given signals based on the mean squared error (MSE) loss function. Finally, created BiLSTM layers are fed with input feature sequences and named according to input features such RAW-Net, STFT-Net, CWT-Net, and WSST-Net.

\subsection{Model evaluation}

To evaluate the trained models' performance, EEG and EOG segments were randomly chosen and grouped in training, validation, and test sets with ratios of 80\%, 10\%, and 10\%, respectively. A total of 75,000 ground truth clean EEG segments and 75,000 semi-synthetic contaminated EEG segments which were generated at 15 different SNR levels were divided into three groups. As a result, 120,000 EEG segments were reserved for training, and 15,000 for both validation and test. There are several performance metrics are utilized to evaluate the regression performance of DL models. The most used two qualitative metrics, namely MSE and root mean square error (RMSE) were calculated to analyze models' robustness. MSE is a measure of the prediction capability of a model. The RMSE is used to assess the error of predictions. Both can be calculated as in equation (\ref{equ3}):
\begin{equation}
RMSE=\sqrt{MSE}=\sqrt{\frac{\sum\limits_{i=1}^{n}(EEG_{i}-denoisedEEG_{i})^{2}}{n}}
\label{equ3}
\end{equation}

\section{Results and Discussion}

The structured BiLSTM-based four different DL models were trained with generated semi-synthetic data by using an NVIDIA 3080 Ti GPU in the MATLAB Deep Learning Toolbox environment. For all architecture, while both input sequence layer and fully connected layer output feature count was set to 1 and both forward and backward LSTM layers' hidden units parameter was set to 150, input features counts are variable due to the TFA methods specialty (\textit{see Table \ref{table1}}). Therefore the trainable parameters and elapsed times (ET) for the training of the BiLSTM networks are also variable. Also, the initial learning rate, epoch, and batch size were adjusted as 0.001, 10, and 150, respectively. Adam optimizer was also utilized in the training of architectures.  

\begin{table}[b]
\vspace{-6mm}
\centering
\caption{Performance Comparison of Proposed Models. }
\begin{tabular}{ccccc} 
\hline
\textbf{Model Name} & \textbf{Feature} & \textbf{Val. RMSE} & \textbf{Avg. MSE} & \textbf{ET (min)}  \\ 
\hline
No Denoising         &    N/A &        N/A              &        1.0244        &         N/A             \\ 

RAW-Net              &  1x512  &          69.242            &        0.5575        &   $\sim$29                \\ 

STFT-Net             &   130x385  &          6.268          &        0.4510        &          $\sim$109                \\ 

CWT-Net              &  94x512  &         0.7148             &        0.4243        &            $\sim$148            \\ 

WSST-Net             &  512x512  &       0.1218               &        0.3066        &            $\sim$308            \\
\hline
\label{table1}
\end{tabular}
\end{table}

The BiLSTM model was first trained using the raw EEG segments without any feature extraction in order to determine a comparable base for TF-based feature extraction methods. Also, the average MSE value of all EEG segments was calculated as 1.0244 to evaluate the denoising performances. Each trained raw segment contains 1x512 data due to each segment was 2 s-long sizes and the sampling rate was 256. For training the RAW-Net, validation RMSE was found as 69.242 with a training time of nearly 29 min. In Table \ref{table1}, average MSE values were calculated by using all test data with different SNR levels at -10dB to +4dB. The RAW-Net's final score was 0.5575. In the STFT-Net model, STFT was computed with a 128-point size regular rectangular window and the overlap size was set to 127 in order to yield the best TF resolution. As mentioned before, only the real part of TF coefficients was considered by stacking the real component on top of the imaginary component for the training of TF-based BiLSTM models due to the LSTM layers cannot deal with complex inputs. As a result, a total of 130x385 real features are extracted from each EEG segment for STFT-Net. Also, the inverse transform of STFT (ISTFT) was also calculated for all predicted outputs provided from the regression layer. Because, when the BiLSTM architecture's inputs are TF coefficients, the output regression layer provides predicted TF coefficients and the main goal is to reach the predicted time-series signal. Hence, the inverse transform of all TF-based methods was also calculated for predicted outputs. The STFT-Net has yielded a 6.268 RMSE value which is nearly 11 times better than RAW-Net. In the CWT-Net, TF feature extraction was performed by using the ``Bump Wavelet''. After computation 94x512 real feature matrix was obtained for each segment. Similarly, the inverse CWT (ICWT) was computed for the performance calculation. As a result, CWT-Net significantly improves the denoising performance with a 0.7148 validation RMSE. Finally, the WSST-Net was trained with real parts of WSST coefficients by using the ``Amor Wavelet''. A total of 512x512 TF coefficient was calculated for each segment and validation RMSE reached 0.1218 which was computed by using the inverse WSST (IWSST). Although the WSST-Net increased the ET for training, it also significantly improve the regression performance.  

\begin{figure}[!htp]
\vspace{-3mm}
\centering
\includegraphics[width=0.8\columnwidth]{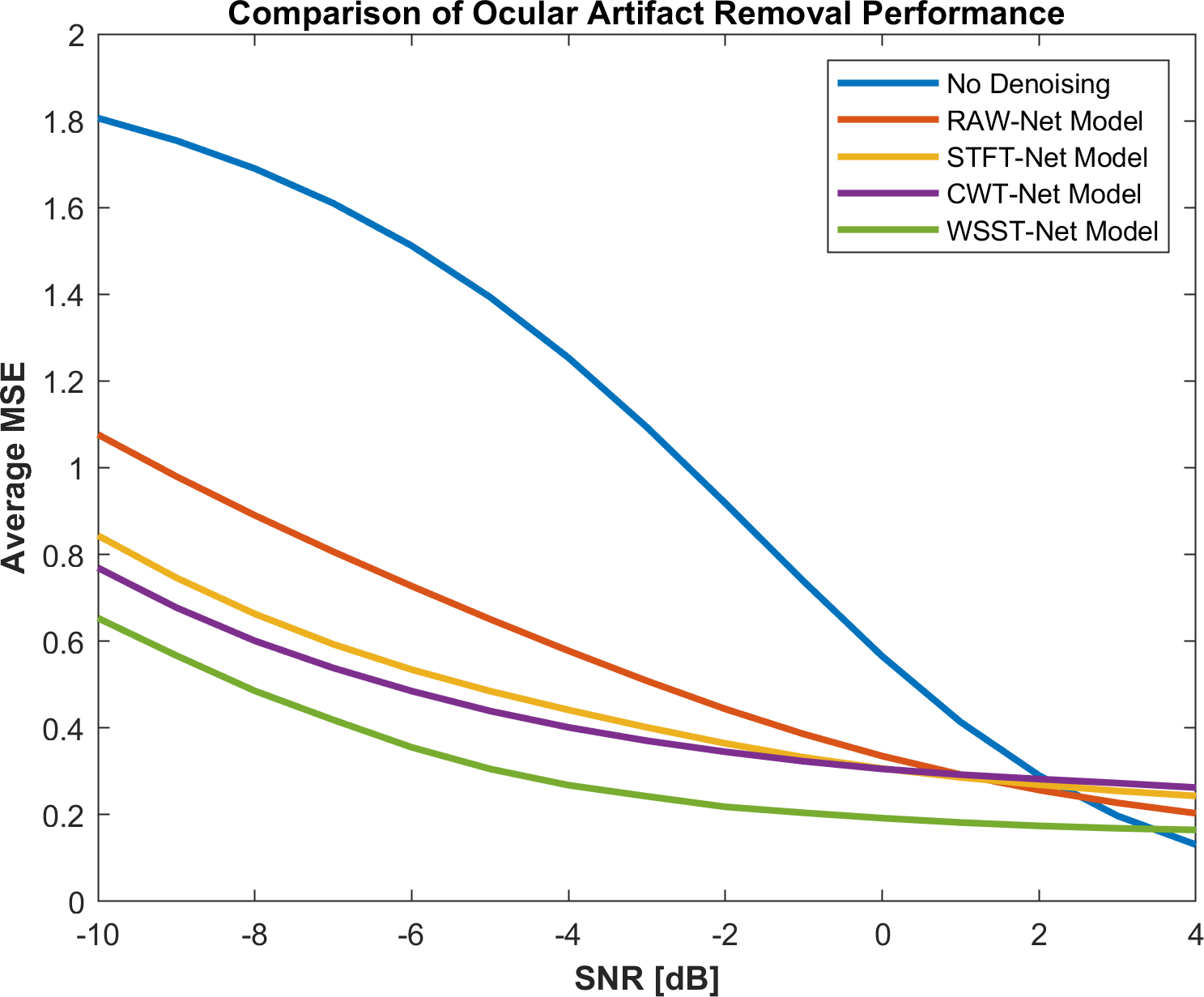}
\caption{Ocular artifact denoising performances of BiLSTM-based proposed models at different SNR levels.}
\label{Fig2}
\vspace{-2mm}
\end{figure}

\begin{figure*}[!ht]
\vspace{-2mm}
\centering
\includegraphics[width=0.7\textwidth]{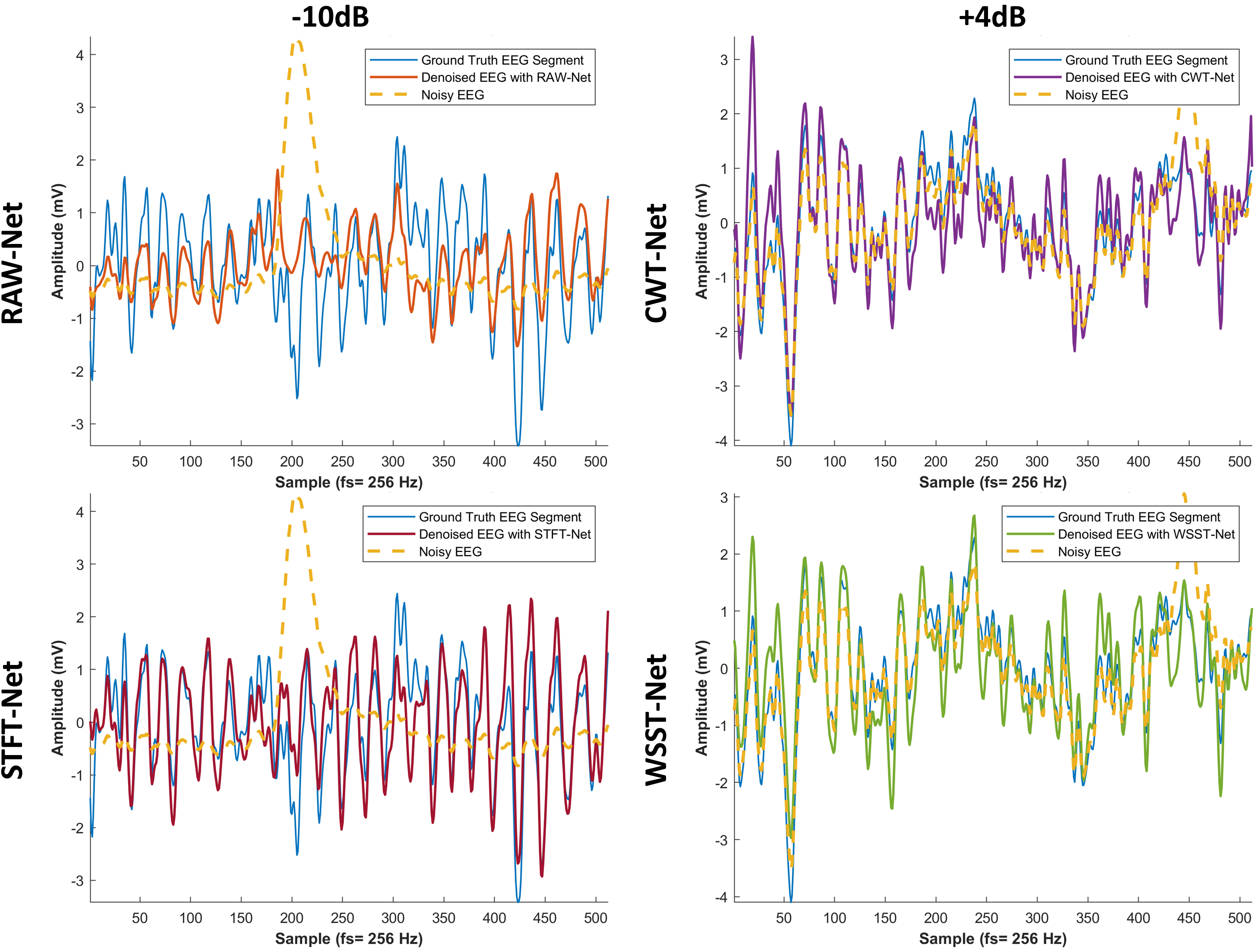}
\caption{Examples of different predictive results of proposed models at -10dB and 4dB.}
\label{Fig3}
\end{figure*}

Fig. \ref{Fig2} shows a comparison of the average MSE values obtained in the test phases of proposed models at different SNR levels. At SNR values, a total of 1000 segments that contain equal clean and contaminated EEG segment distribution were predicted with the proposed models. Thus, each point at the SNR levels indicates the average MSE result of those segments. Also, the average MSE of un-denoised segments was provided for benchmarking the proposed models. When obtained results at higher SNR values around 4dB are examined, almost all outperformed. Contrary, at the lowest SNRs, the performance distinction stands out. Also, when the lower performance of the Raw-Net model is considered, TF-based feature extraction methods significantly improve the prediction performance. Moreover, WSST-Net outperformed at all SNRs than other models. Additionally, all models have lower success at the lowest SNRs than at higher SNRs. All models had smooth MSE decreasing proportionally to SNR increasing.

\begin{table*}[ht]
\begin{center}
\caption{Comparison of State-of-the-art Ocular Artifact Removal Approaches.}
\label{table2}
\begin{tabular}{ccccccc}
\toprule
\textbf{Work}  & \textbf{Dataset} & \textbf{Length} & \textbf{Size}& \textbf{Feature}& \textbf{Arch.}  & \textbf{RMSE} 
\\ \toprule

\multirow{2}{*}{Nguyen et al.{}\cite{nguyen2012eog19}}& Driving Test  & \multirow{2}{*}{30 s} & \multirow{2}{*}{N/A}  & \multirow{2}{*}{WD} & \multirow{2}{*}{WNN} & 12.2473       \\ 
{} & Visual Selection Task & {}  & {}  & {} & {} & 19.2154 \\ \cmidrule{1-7}

Yang et al.\cite{yang2018automatic20} & BCI Competition IV    & 5 s  & 31,978 for each subject   & CSP & SSAE & 1.2838        \\ \cmidrule{1-7}

Sun et al.\cite{sun2020novel21}{}   & CHB-MIT Scalp EEG & N/A  & 921,600 for each subject   & N/A & 1D-ResCNN & 9.83   \\ \cmidrule{1-7}

Mashhadi et al.{{}\cite{mashhadi2020deep23}{} }  & Semi-Simulated EEG & 30 s   & N/A    & N/A    & U-Net   & 0.0239  \\ \cmidrule{1-7}

Lopes et al.{{}\cite{lopes2021automatic24}{} }   & EPILEPSIAE & 10 min    & 4309 EEG  & ICA & DCNN  & 4.8 \\ \cmidrule{1-7}

\multirow{3}{*}{Zhang et al.{}\cite{zhang2021eegdenoisenet26}}  &\multirow{3}{*} {EEGdenoiseNet}   &\multirow{3}{*}{2 s}  & \multirowcell{3}{ {4514 clean EEG}\\ {3400 EOG}} & \multirow{3}{*}{N/A} & FCNN & {0.2221} \\

{} & {}& {}& {}& {}& Simple CNN & {0.1448} \\ 
{} & {}& {}& {}& {}& LSTM& {0.1888} \\  \cmidrule{1-7}

\multirow{4}{*}{\textbf{This Work}} & \multirowcell{4}{\textbf{EEGdenoiseNet} \\\textbf{+}\\ \textbf{DEAP}} & \multirow{4}{*}{\textbf{2 s}}  & \multirowcell{4}{\textbf{6164 clean EEG} \\\textbf{5000 EOG} \\\textbf{150,000 augmented}}  & \textbf{N/A}    & \multirow{4}{*}{\textbf{BiLSTM}}   & \textbf{0.7466}       \\   

{} & {}& {}& {}& {\textbf{STFT}} & {} & \textbf{0.6716}  \\ 
{} & {}& {}& {}& {\textbf{CWT}}& {} & \textbf{0.6513} \\ 
{} & {}& {}& {}& {\textbf{WSST}}& {} &  \textbf{0.5537} \\ \bottomrule
\end{tabular}
\end{center}
\vspace{-3mm}
\end{table*}

Fig. \ref{Fig3} illustrates four examples of prediction outputs for 2 s-long segments at two different SNR values. Ground truth clean EEG and contaminated EEG segments were also provided. All models' outcomes revealed that the contaminated EEG was nearly denoised. Considering the performance of the Raw-Net and STFT-Net models at the lowest SNR value of -10 dB for the same segment, the STFT-Net outcome creates a denoised EEG segment significantly similar to the ground truth. Furthermore, at 4dB SNRs that are more easily cleanable contaminated segments, this performance distinction was not significant for both CWT-Net and WSST-Net. However, both superiorly provided predictive segments which were similar to the ground truth EEG segment.

The obtained results of this study were also compared to state-of-the-art studies recently reported in the literature given in Table \ref{table2}. Almost all are focused on different datasets, segment sizes, and lengths, as well as feature extraction methods and DL-based architectures. Their main goal was to reach as lowest as possible RMSE values. Our proposed models demonstrated that their performance has a significant impact according to RMSE values. Also, we proposed a novel feature extraction method based on WSST for ocular artifact removal, and a comprehensive comparison is performed to reveal the enhancement. According to Table \ref{table2}, the proposed model outperformed most of the others. Among them had significant complex approaches as well as higher computational costs. Additionally, the most utilized approach, semi-synthetic noisy EEG data generation was limited a narrow SNR bands. Contrary, we evaluate 15 different SNR levels for different BiLSTM-based models. Also, we evaluate the model performances by combining different datasets to provide an end-to-end DL model for generalization. Moreover, our results provided the utilization of TFA methods for generating time-series frequency components by easy adaptation of the real part of TF coefficients. Hence, a low-cost approach that only depends on TFA complexity and DL network was proposed and the adaptability of TFA methods for ocular artifact removal was demonstrated. Finally, a novel BiLSTM-based DL architecture that contains a few layers was proposed to effectively remove EOG in contaminated EEG signals. Our proposed method emphasized a low-cost approach compared to existing studies in the literature, the superior success of the LSTM-regression architecture, and the adaptability of TFA methods in the prediction of time-series biological signals such as EEG with LSTM architectures.

\section{Conclusion}

This study presents an effective and novel ocular artifact removal approach from EEG signals developed on novel BiLSTM-regression architecture and the WSST method. For ocular artifact removal, the highly localized and high-resolution WSST coefficients are utilized to feed the BiLSTM network, for the first time. The semi-synthetic data was generated at wide SNR levels to evaluate the performance of trained models. The experimental results demonstrated that the proposed WSST-based BiLSTM method yielded superior performance compared to both traditional methods, STFT-Net and CWT-Net, and similar approaches in the literature. Furthermore, the WSST-Net model is encouraging success at lower SNR levels which is almost as well as at higher SNR levels. Obtained results highlighted the comparability to ICA and encouraged the possibility of automated removal of ocular artifacts by a DL framework. Finally, the proposed approach can easily be adapted to the EEGLAB toolbox and could be expanded to remove other artifacts, such as muscle artifacts.





\bibliographystyle{IEEEtran}

\bibliography{references.bib}

\begin{thebibliography}{10}
\providecommand{\url}[1]{#1}
\csname url@samestyle\endcsname
\providecommand{\newblock}{\relax}
\providecommand{\bibinfo}[2]{#2}
\providecommand{\BIBentrySTDinterwordspacing}{\spaceskip=0pt\relax}
\providecommand{\BIBentryALTinterwordstretchfactor}{4}
\providecommand{\BIBentryALTinterwordspacing}{\spaceskip=\fontdimen2\font plus
\BIBentryALTinterwordstretchfactor\fontdimen3\font minus
  \fontdimen4\font\relax}
\providecommand{\BIBforeignlanguage}[2]{{%
\expandafter\ifx\csname l@#1\endcsname\relax
\typeout{** WARNING: IEEEtran.bst: No hyphenation pattern has been}%
\typeout{** loaded for the language `#1'. Using the pattern for}%
\typeout{** the default language instead.}%
\else
\language=\csname l@#1\endcsname
\fi
#2}}
\providecommand{\BIBdecl}{\relax}
\BIBdecl

\bibitem{teplan2002fundamentals}
M.~Teplan \emph{et~al.}, ``Fundamentals of eeg measurement,'' \emph{Measurement
  science review}, vol.~2, no.~2, pp. 1--11, 2002.

\bibitem{mannan2018identification}
M.~M.~N. Mannan, M.~A. Kamran, and M.~Y. Jeong, ``Identification and removal of
  physiological artifacts from electroencephalogram signals: A review,''
  \emph{Ieee Access}, vol.~6, pp. 30\,630--30\,652, 2018.

\bibitem{sweeney2012artifact}
K.~T. Sweeney, T.~E. Ward, and S.~F. McLoone, ``Artifact removal in
  physiological signals-practices and possibilities,'' \emph{IEEE transactions
  on information technology in biomedicine}, vol.~16, no.~3, pp. 488--500,
  2012.

\bibitem{ranjan2021ocular}
R.~Ranjan, B.~C. Sahana, and A.~K. Bhandari, ``Ocular artifact elimination from
  electroencephalography signals: A systematic review,'' \emph{Biocybernetics
  and Biomedical Engineering}, vol.~41, no.~3, pp. 960--996, 2021.

\bibitem{ccinar2017novel}
S.~Cinar and N.~Acir, ``A novel system for automatic removal of ocular
  artefacts in eeg by using outlier detection methods and independent component
  analysis,'' \emph{Expert Systems with Applications}, vol.~68, pp. 36--44,
  2017.

\bibitem{minguillon2017trendseeg}
J.~Minguillon, M.~A. Lopez-Gordo, and F.~Pelayo, ``Trends in eeg-bci for
  daily-life: Requirements for artifact removal,'' \emph{Biomedical Signal
  Processing and Control}, vol.~31, pp. 407--418, 2017.

\bibitem{gorjan2022removaleeg}
D.~Gorjan, K.~Gramann, K.~D. Pauw, and U.~Marusic, ``Removal of
  movement-induced {EEG} artifacts: current state of the art and guidelines,''
  \emph{Journal of Neural Engineering}, vol.~19, no.~1, p. 011004, 2022.

\bibitem{barua2014review}
S.~Barua and S.~Begum, ``A review on machine learning algorithms in handling
  eeg artifacts,'' in \emph{The Swedish AI Society (SAIS) Workshop, 14, 2014,
  Stockholm, Sweden}, 2014.

\bibitem{lawhern2013detect}
V.~Lawhern, W.~D. Hairston, and K.~Robbins, ``Detect: A matlab toolbox for
  event detection and identification in time series, with applications to
  artifact detection in eeg signals,'' \emph{PloS one}, vol.~8, no.~4, p.
  e62944, 2013.

\bibitem{vijila2007artifacts}
C.~K.~S. Vijila, P.~Kanagasabapathy, S.~Johnson, and V.~Ewards, ``Artifacts
  removal in eeg signal using adaptive neuro fuzzy inference system,'' in
  \emph{2007 International Conference on Signal Processing, Communications and
  Networking}.\hskip 1em plus 0.5em minus 0.4em\relax IEEE, 2007, pp. 589--591.

\bibitem{schetinin2007bayesian}
V.~Schetinin and C.~Maple, ``A bayesian model averaging methodology for
  detecting eeg artifacts,'' in \emph{2007 15th International Conference on
  Digital Signal Processing}.\hskip 1em plus 0.5em minus 0.4em\relax IEEE,
  2007, pp. 499--502.

\bibitem{fairley2010hybrid}
J.~Fairley, G.~Georgoulas, C.~Stylios, and D.~Rye, ``A hybrid approach for
  artifact detection in eeg data,'' in \emph{International Conference on
  Artificial Neural Networks}.\hskip 1em plus 0.5em minus 0.4em\relax Springer,
  2010, pp. 436--441.

\bibitem{nguyen2012eog19}
H.-A.~T. Nguyen, J.~Musson, F.~Li, W.~Wang, G.~Zhang, R.~Xu, C.~Richey,
  T.~Schnell, F.~D. McKenzie, and J.~Li, ``Eog artifact removal using a wavelet
  neural network,'' \emph{Neurocomputing}, vol.~97, pp. 374--389, 2012.

\bibitem{yang2018automatic20}
B.~Yang, K.~Duan, C.~Fan, C.~Hu, and J.~Wang, ``Automatic ocular artifacts
  removal in eeg using deep learning,'' \emph{Biomedical Signal Processing and
  Control}, vol.~43, pp. 148--158, 2018.

\bibitem{sun2020novel21}
W.~Sun, Y.~Su, X.~Wu, and X.~Wu, ``A novel end-to-end 1d-rescnn model to remove
  artifact from eeg signals,'' \emph{Neurocomputing}, vol. 404, pp. 108--121,
  2020.

\bibitem{mashhadi2020deep23}
N.~Mashhadi, A.~Z. Khuzani, M.~Heidari, and D.~Khaledyan, ``Deep learning
  denoising for eog artifacts removal from eeg signals,'' in \emph{2020 IEEE
  Global Humanitarian Technology Conference (GHTC)}.\hskip 1em plus 0.5em minus
  0.4em\relax IEEE, 2020, pp. 1--6.

\bibitem{lopes2021automatic24}
F.~Lopes, A.~Leal, J.~Medeiros, M.~F. Pinto, A.~Dourado, M.~D{\"u}mpelmann, and
  C.~Teixeira, ``Automatic electroencephalogram artifact removal using deep
  convolutional neural networks,'' \emph{IEEE Access}, vol.~9, pp.
  149\,955--149\,970, 2021.

\bibitem{zhang2021eegdenoisenet26}
H.~Zhang, M.~Zhao, C.~Wei, D.~Mantini, Z.~Li, and Q.~Liu, ``Eegdenoisenet: A
  benchmark dataset for deep learning solutions of eeg denoising,''
  \emph{Journal of Neural Engineering}, vol.~18, no.~5, p. 056057, 2021.

\bibitem{koelstra2011deap}
S.~Koelstra, C.~Muhl, M.~Soleymani, J.-S. Lee, A.~Yazdani, T.~Ebrahimi, T.~Pun,
  A.~Nijholt, and I.~Patras, ``Deap: A database for emotion analysis; using
  physiological signals,'' \emph{IEEE transactions on affective computing},
  vol.~3, no.~1, pp. 18--31, 2011.

\bibitem{makifAbnormal2021}
M.~A. Ozdemir, D.~H. Kisa, O.~Guren, A.~Onan, and A.~Akan, ``Emg based hand
  gesture recognition using deep learning,'' in \emph{2020 Medical Technologies
  Congress (TIPTEKNO)}, 2020, pp. 1--4.

\bibitem{DELORME20049}
A.~Delorme and S.~Makeig, ``Eeglab: an open source toolbox for analysis of
  single-trial eeg dynamics including independent component analysis,''
  \emph{Journal of Neuroscience Methods}, vol. 134, no.~1, pp. 9--21, 2004.

\bibitem{wang2015removal}
G.~Wang, C.~Teng, K.~Li, Z.~Zhang, and X.~Yan, ``The removal of eog artifacts
  from eeg signals using independent component analysis and multivariate
  empirical mode decomposition,'' \emph{IEEE journal of biomedical and health
  informatics}, vol.~20, no.~5, pp. 1301--1308, 2015.

\bibitem{ozdemir2021EEG}
M.~A. Ozdemir, O.~K. Cura, and A.~Akan, ``Epileptic eeg classification by using
  time-frequency images for deep learning,'' \emph{International Journal of
  Neural Systems}, vol.~31, no.~08, p. 2150026, 2021.

\bibitem{sadowsky1996investigation}
J.~Sadowsky, ``Investigation of signal characteristics using the continuous
  wavelet transform,'' \emph{johns hopkins apl technical digest}, vol.~17,
  no.~3, pp. 258--269, 1996.

\bibitem{vetterli1991wavelets}
M.~Vetterli, ``Wavelets and signal processing,'' \emph{IEEE ASSP Magazine},
  vol.~8, 1991.

\bibitem{oppenheim1998spectrum}
A.~Oppenheim, R.~Schafer, and J.~Buck, ``Spectrum analysis of random signals
  using estimates of the autocorrelation sequence,'' \emph{Discrete-Time Signal
  Processing, Prentice-Hall, New Jersey}, vol. 7458, 1998.

\bibitem{kumar2020improved}
A.~Kumar, C.~Gandhi, Y.~Zhou, G.~Vashishtha, R.~Kumar, and J.~Xiang, ``Improved
  cnn for the diagnosis of engine defects of 2-wheeler vehicle using wavelet
  synchro-squeezed transform (wsst),'' \emph{Knowledge-Based Systems}, vol.
  208, p. 106453, 2020.

\bibitem{rahman2022extended}
A.~U. Rahman, A.~Tubaishat, F.~Al-Obeidat, Z.~Halim, M.~Tahir, and F.~Qayum,
  ``Extended ica and m-csp with bilstm towards improved classification of eeg
  signals,'' \emph{Soft Computing}, pp. 1--12, 2022.

\end{thebibliography}

\end{document}